\documentclass[a4paper,11pt]{article}

\usepackage{jcappub} 
\usepackage[T1]{fontenc}

\usepackage{amsmath}
\usepackage{amsfonts}
\usepackage{amssymb}
\usepackage{mathrsfs}

\usepackage{graphicx}
\usepackage{color}
\usepackage{soul}
\usepackage{subfig}
\usepackage[normalem]{ulem}
\usepackage[utf8]{inputenc}
\usepackage{aas_macros}

\usepackage[T1]{fontenc} 

\renewcommand{\vec}[1]{\boldsymbol{#1}}

\newcommand{\dif}{\mathrm{d}}

\usepackage{floatrow}
\DeclareFloatFont{tiny}{\scriptsize}
\title{\boldmath Evidence of fresh cosmic ray in galactic plane based on DAMPE measurement of B/C and B/O ratios}

\author[a,b,c]{Pei-Pei Zhang,}
\author[a,b]{Xin-Yu He,}
\author[c,1]{Wei Liu}
\author[c,d,1]{and Yi-Qing Guo \note{Corresponding author}}

\affiliation[a]{Key Laboratory of Dark Matter and Space Astronomy, Purple Mountain Observatory, 
Chinese Academy of Sciences, Nanjing 210023, China}
\affiliation[b]{School of Astronomy and Space Science, University of Science and Technology of China, 
Hefei 230026, Anhui, China}
\affiliation[c]{Key Laboratory of Particle Astrophysics, Institute of High Energy Physics, 
Chinese Academy of Sciences, Beijing 100049, China}
\affiliation[d]{University of Chinese Academy of Sciences, Beijing 100049, China}

\emailAdd{ppzhang@pmo.ac.cn}
\emailAdd{xyhe@pmo.ac.cn}
\emailAdd{liuwei@ihep.ac.cn}
\emailAdd{guoyq@ihep.ac.cn}

\abstract{
More and more experiments have identified that the energy spectra of both primary and secondary cosmic-rays exhibit a hardening above $\sim 200$ GV. More recently, the DAMPE experiment has reported a hardening of boron-to-carbon ratio at $200$ GV. These signs call for modifications of the conventional cosmic-ray (CR) picture. In this work, we propose that the plethoric secondary cosmic rays, for example, boron, antiprotons, originate from the hadronic interactions of freshly accelerated cosmic rays with the interstellar gas near the sources. We find that secondary-to-primary ratios, for example, boron-to-carbon, boron-to-oxygen and antiproton-to-proton ratios, can be well described. The measurements of electrons and positrons can also be accounted for.
}

\begin{document}
\maketitle
\flushbottom

\section{Introduction}
\label{sec:intro}

In the standard picture of acceleration and propagation, the cosmic-ray (CR) flux is expected to fall off with energy as a single power-law \citep{2002astro.ph.12111M, 2007ARNPS..57..285S}. The part accelerated within the supernova remnants (SNRs) are usually called primary CRs, such as protons, helium, carbon, oxygen, iron, electron, etc. The energy dependence of the observed flux is $\Phi_p \propto E^{-\alpha -\delta}$, with $\alpha$ and $\delta$ being the power index of the injection and the exponent of the diffusion coefficient respectively. The other part of CRs, i.e., secondary CRs, for example, lithium, beryllium, boron, positrons and antiprotons, are generated by the hadronic interactions between the primary CRs and the interstellar medium (ISM) during transport in the Galaxy. Compared with the primary ones, the energy spectrum of the secondaries is $\Phi_s \propto E^{-\alpha -2\delta}$. Thus, the ratio of secondary-to-primary nuclei is expected to also have a power-law relation against energy, i.e., $\Phi_s/\Phi_p \propto E^{-\delta}$.

However, more and more observations have challenged such traditional impressions. A number of experiments have confirmed that the spectra of primary CR components including proton, helium, carbon, oxygen, neon, magnesium, and silicon, harden at more than $\sim 200$ GV \citep{2007BRASP..71..494P, 2009BRASP..73..564P, 2010ApJ...714L..89A, 2011ApJ...728..122Y, 2017ApJ...839....5Y, 2011Sci...332...69A, 2015PhRvL.114q1103A, 2015PhRvL.115u1101A, PhysRevLett.119.251101, 2020PhRvL.124u1102A}. Recently, both proton and helium spectra have been observed to soften at higher energies($> \sim 10$ TeV) \citep{2017JCAP...07..020A, 2017ApJ...839....5Y, 2019SciA....5.3793A, 2021PhRvL.126t1102A}. The same applies to the secondary components. The measurements of lithium, beryllium, and boron showed that their fluxes harden above $200$ GV, likewise, with an average hardening of $0.13$, larger than helium, carbon, and oxygen. This indicates a hardening of boron-to-carbon ratio at high energies \citep{2022MNRAS.516.3470H}. More recently, the DAMPE experiment has detected an explicit hardening of the boron-to-carbon ratio at $\sim 200$ GV \citep{DAMPECOLLABORATION20222162}. In addition, the measurements have revealed that the antiproton-to-proton ratio does not exhibit an explicit rigidity dependence between $\sim 60$ GV and $\sim 450$ GV \citep{2016PhRvL.117i1103A}. 

As for electrons and positrons, the accurate measurements have also unveiled abundant unexpected features. A remarkable positron excess above $10$ GeV has been identified by a lot of experiments \citep{2009Natur.458..607A, 2009PhRvL.102r1101A, 2019PhRvL.122d1102A}. The latest measurement by the AMS-02 experiment found a sharp drop-off in the positron spectrum starting at $\sim$ 284 GeV \citep{2019PhRvL.122d1102A}. The precise measurements have also revealed a significant excess above $\sim50$ GeV in the electron spectrum \cite{2017Natur.552...63D, 2019PhRvL.122j1101A}. Furthermore, the ground-based telescopes reported a knee-like break around $1$ TeV in the all-electron spectrum \citep{2009A&A...508..561A, 2008PhRvL.101z1104A, hess-icrc2017, 2011ICRC....6...47B, 2018PhRvD..98f2004A}. Later, this feature was verified by the space-borne experiment \citep{2017Natur.552...63D}.

The findings of spectral hardening of nuclei bring about various alternatives of the traditional CR theory. Most of them fall into, but not limited to, three categories: acceleration process \citep{2011ApJ...729L..13O, 2012PhRvL.108h1104M, 2010ApJ...725..184B, 2011PhRvD..84d3002Y, 2017ApJ...835..229K}, transport effect \citep{2012PhRvL.109f1101B, 2012ApJ...752L..13T, 2012JCAP...01..010B, 2012ApJ...752...68V, 2014A&A...567A..33T, 2014ApJ...782...36E, 2015A&A...583A..95A, 2015PhRvD..92h1301T, 2016ApJ...819...54G,2016ApJ...827..119C,2016PhRvD..94l3007F, 2016ApJ...819...54G, 2016ChPhC..40a5101J, 2017PhRvD..95h3007Y, 2018ChPhC..42g5103G, 2018PhRvD..97f3008G,2022arXiv221009205M}, and nearby source(s) \citep{2012A&A...544A..92B, 2012MNRAS.421.1209T, 2013MNRAS.435.2532T, 2013A&A...555A..48B, 2015RAA....15...15L, 2015ApJ...803L..15T, 2015ApJ...815L...1T, 2017PhRvD..96b3006L,2018PhRvL.120d1103K, 2020FrPhy..1524601Y, 2021FrPhy..1624501Y, PhysRevD.105.023002}. For transport effect and close-by sources, an excess of boron-to-carbon ratio is also expected in these models. 
The discovery of a positron excess with respect to the conventional model implies the existence of extra primary components. They are attributed to either astrophysical, such as local pulsars \citep{1970ApJ...162L.181S, 2001A&A...368.1063Z, 2009PhRvL.103e1101Y, 2009JCAP...01..025H, 2015APh....60....1Y, 2012CEJPh..10....1P, 2016ApJ...819...54G} and the hadronic interactions inside supernova remnants (SNRs) \citep{2009PhRvL.103e1104B, 2009PhRvD..80f3003F, 2009ApJ...700L.170H, 2015ApJ...803L..15T, 2017PhRvD..96b3006L}, or more exotic origins like the dark matter self-annihilation or decay \citep{2008PhRvD..78j3520B,2009NuPhB.813....1C, 2009PhLB..672..141B, 2009PhRvD..79b3512Y, 2009PhRvL.103c1103B, 2009PhRvD..80b3007Z}.


Recently, the HAWC collaboration has detected the extended TeV gamma-ray emission of two middle-aged pulsars, Geminga and PSR B0656+14 \citep{2017Sci...358..911A}, and the LHAASO collaboration has measured the energy spectrum of gamma-rays surrounding pulsar J0621+3755 \citep{2021PhRvL.126x1103A}. By fitting the gamma-ray profile using the diffusion model, they found that the diffusion length near the source is much smaller than that in the Galaxy, which is derived by fitting the B/C ratio. This means that before diffusion in the interstellar space, the cosmic rays experience a slower diffusion process near the source region \citep{2016ChPhC..40k5001G, PhysRevD.100.063020}, which has not been considered in the conventional propagation model. 
In this work, we consider the scenario where secondary CRs generate from the hadronic interactions between freshly accelerated cosmic rays and the medium when CRs experience a slower diffusion near the sources. 
{This is similar with the nested leaky box model \citep{2016ApJ...827..119C} in which a two-stage leakage of particles in the vicinities of the sources (cocoons) and the Milky Way is assumed.
Our scenario is a more physical realization of the simplified nested leaky box model. }
We find that the ratio anomalies, i.e., boron-to-carbon, boron-to-oxygen and antiproton-to-proton ratios can be naturally accounted for. The excess of positrons can also be explained. We further find that above TeV energies, the electron and positron spectra give rise to extra hardening. This could be testified by the future experiments.

\section{Model description}

\subsection{Spatial-dependent propagation}
After leaving the acceleration sites, CRs perform a diffusive motion within a huge region, commonly referred to as the diffusive halo. The general transport equation is written as
\begin{eqnarray}
\frac{\partial \psi}{\partial t} &=& Q(\vec{r}, p) + \nabla \cdot ( D_{xx}\nabla\psi - \vec{V}_{c}\psi )
+ \frac{\partial}{\partial p}\left[p^2D_{pp}\frac{\partial}{\partial p}\frac{\psi}{p^2}\right]
\nonumber\\
&& - \frac{\partial}{\partial p}\left[ \dot{p}\psi - \frac{p}{3}(\nabla\cdot\vec{V}_c)\psi \right]
- \frac{\psi}{\tau_f} - \frac{\psi}{\tau_r} ~,
\label{propagation_equation}
\end{eqnarray}
where $\psi = \dif n/\dif p$ is the CR density per total particle momentum $p$ at position $\vec{r}$. In this work, the spatially-dependent propagation (SDP) is introduced to work out the distribution of CRs in the Galaxy. For a comprehensive introduction to SDP model, one can refer to \cite{2012ApJ...752L..13T,2016ApJ...819...54G,2018ApJ...869..176L}. 

In the SDP model, the entire diffusive halo is split into two regions characterized according to their diffusion properties. In the Galactic disk and its surrounding areas, i.e., the so-called inner halo (IH), the turbulence is intense due to the activities of supernova explosions, and correspondingly, the diffusion process is slow. Outside the IH, the turbulence is driven by CRs themselves, so the diffusion process tends to be fast, approaching the conventional diffusion model. These areas are called the outer halo (OH). The half-thickness of the whole diffusive halo is $L$, with the fractions of IH and OH being $\xi$ and $1-\xi$, respectively. At the halo border, it is imposed that $\psi(R, z, p) =  \psi(r, \pm z_h, p) = 0$, namely, the free-escape condition. The diffusion coefficient $D$ in the whole region is parameterized as \citep{2016ApJ...819...54G, 2018ApJ...869..176L}:
\begin{equation}
D(r,z, {\cal R} )= D_{0}F(r,z)\beta^{\eta} \left(\dfrac{\cal R}{{\cal R}_{0}} \right)^{\delta_0 F(r,z)} ~,
\label{eq:diffusion}
\end{equation}
with
\begin{equation}
F(r,z) =
\begin{cases}
g(r,z) +\left[1-g(r,z) \right] \left(\dfrac{z}{\xi L} \right)^{n} , &  |z| \leqslant \xi L \\
1 ~, & |z| > \xi L
\end{cases} ~.
\end{equation}
In the IH, the turbulence level is expected to be positively associated with the large population of sources, and thus the diffusion coefficient is anti-correlated with the source
density. Here, $g(r,z)$ is $N_m/[1+f(r,z)]$, in which $f(r,z)$ is the source distribution.

\subsection{Source Term}

\subsubsection{Primary CRs}
The spatial distribution of SNRs is approximately parameterized as an axisymmetric function:
\begin{equation}
f(r, z) = \left(\dfrac{r}{r_\odot} \right)^\alpha \exp \left[-\dfrac{\beta(r-r_\odot)}{r_\odot} \right] \exp \left(-\dfrac{|z|}{z_s} \right) ~,
\label{eq:radial_dis}
\end{equation}
in which $r_\odot = 8.5$ kpc is the distance of the solar system from the Galactic center. The values of parameters $\alpha$ and $\beta$, taken from \citep{1996A&AS..120C.437C}, are $1.69$ and $3.33$, respectively. Away from the Galactic plane, the SNR density decreases exponentially, with a characteristic height of $z_{s} = 200$ pc.

The injection spectrum of primary CRs in each source $Q(p)$ is assumed to be rigidity-dependent and follows a broken power-law, i.e.,
\begin{equation}
  q(\mathcal R) =  q_0 \times\left\{ \begin{array}{ll}
    \left( \dfrac{\mathcal R}{\mathcal R_{\rm br}} \right)^{-\nu_1} ~, & \mathcal R \leqslant \mathcal R_{\rm br}\\
    \left( \dfrac{\mathcal R}{\mathcal R_{\rm br}} \right)^{-\nu_2} \exp\left[-\dfrac{\mathcal R}{\mathcal R_{\rm c}} \right] ~, & \mathcal R > \mathcal R_{\rm br}
  \end{array}
  \right. ~,
\label{inject_spec_nuclei}
\end{equation}
where $q_0$ is the normalization factor; $\nu_1$($\nu_2$) is the spectral index below(above) $\mathcal R_{\rm br}$; $\mathcal R_{\rm c}$ is the cutoff rigidity.

\subsubsection{Secondary CRs}
Secondary particles, such as lithium, beryllium, boron and antiprotons, are brought forth throughout the transport by spallation and radioactive decay. To account for the production of lithium, beryllium and boron, the so-called straight-ahead approximation is adequate, in which the kinetic energy per nucleon is conserved during the interactions. Their source terms read
\begin{equation}
Q_{j = \rm Li, Be, B} = \sum_{i = \rm C, N, O} (n_{\rm H} \sigma_{i+{\rm H}\rightarrow j} +n_{\rm He} \sigma_{i+{\rm He} \rightarrow j} ) v \psi_i ~,
\end{equation}
where $n_{\rm H/He}$ is the number density of the interstellar hydrogen/helium and $\sigma_{i+{\rm H/He} \rightarrow j}$ is the total cross section of the corresponding hadronic interactions. Unlike the light nuclei mentioned above, the generation of antiprotons is expressed as a convolution of the primary spectra $\psi_i(p)$ and the relevant differential cross section $d \sigma_{i + {\rm H/He} \to j}/d E_j$, i.e.,
 \begin{eqnarray}
\nonumber Q_j &=& \sum_{i = \rm p, He} \int dp_i v \left\lbrace n_{\rm H} \frac{\dif \sigma_{i+{\rm H}\rightarrow j}(p_i, p_j)}{\dif p_j} +n_{\rm He} \frac{\dif \sigma_{i+{\rm He}\rightarrow j}(p_i, p_j)}{\dif p_j} \right\rbrace  \psi_i(p_i) ~.
\end{eqnarray}

Furthermore, the CRs freshly accelerated in the sources could interact with the gas around the sources before they enter the interstellar space. The injection spectra of lithium, beryllium and boron near the sources are written as
\begin{eqnarray}
Q_{j={\rm Li, Be, B}} = \sum_{i = \rm C, N, O} (n_{\rm H} \sigma_{i+{\rm H}\rightarrow j} +n_{\rm He} \sigma_{i+{\rm He}\rightarrow j} )v Q_i(E). 
\end{eqnarray}
and those of positrons and antiprotons are written as
\begin{equation}
\begin{split}
Q_{j}  \; = \;
\sum_{i = \rm p, He} {\displaystyle \int\limits_{E_{\rm th}}^{+ \infty}} \; d E_i \;
v \; \left\lbrace n_{\rm H}
{\displaystyle \frac{d \sigma_{i + {\rm H} \to j}}{d E_j }} 
+n_{\rm He} {\displaystyle \frac{d \sigma_{i + {\rm He} \to j}}{d E_j }} \right\rbrace
Q_i(E_i) ~.
\label{sec_source}
\end{split}
\end{equation}

\begin{table*}
\centering
\caption{The parameters of the transport equations.}
\begin{tabular}{ccccccccc}
\hline
\hline
  ${D}_0~[\rm cm^{-2}s^{-1}]$ & $\delta_0$  & ${N}_{m}$  & $\xi$  & $n$ & $\eta$ & ${\cal R}_0~$ & ${\cal \nu}_{\rm A}$~[km s$^{-1}$]   & ${z}_h$~[kpc] \\
\hline
  $4.87$ & $0.65$  &  $0.57$ & $0.1$ &  $4.0$  & $0.05$  & $4$  &$6$ & $5$ \\
\hline
\hline
\end{tabular}
\label{table_1}
\end{table*}

\begin{table*}
\centering
\caption{Fitting parameters of the injection spectra.}
\begin{tabular}{|c|c c c c c|c c c|c c c|}

\hline
  & & & background & & & & local&\\
\hline
  Element &  ${Q}_0~ [\rm m^{-2}sr^{-1}s^{-1}GeV^{-1}]^\dagger$ & $\nu1$ & $\nu2$ & ${\cal R}_{\rm br}$ [GV] & ${\cal R}_{\rm c}$ [TV]  & $q_0~ [\rm GeV^{-1}]$  
  & $\gamma$ &  ${\cal R}_{\rm c}$ ~[TV]\\
\hline
  e    & $2.7\times 10^{-1}$ & $1.84$ & $2.73$  & $5.0$  &  $10$ & $5.5\times 10^{49}$ &  $2.1$  & $28$ \\
  p    & $3.65\times 10^{-2}$ & $1.41$ & $2.35$  & $7.6$ &  $3000$ & $4.9\times 10^{52}$ &  $2.1$  & $21$ \\
  C    & $8.64\times 10^{-5}$ & $1.00$ & $2.27$  & $1.9$ &   $3000$ & $5.0\times 10^{50}$ &  $2.1$  & $21$ \\
  O    & $1.06\times 10^{-4}$ & $1.00$ & $2.27$  & $1.9$ &   $3000$ & $4.0\times 10^{50}$ &  $2.1$  & $21$ \\
\hline
\end{tabular}
\label{table-parm}
\end{table*}

\section{Results}

\begin{figure*}[!htb]
\centering
\includegraphics[height=6.cm, angle=0]{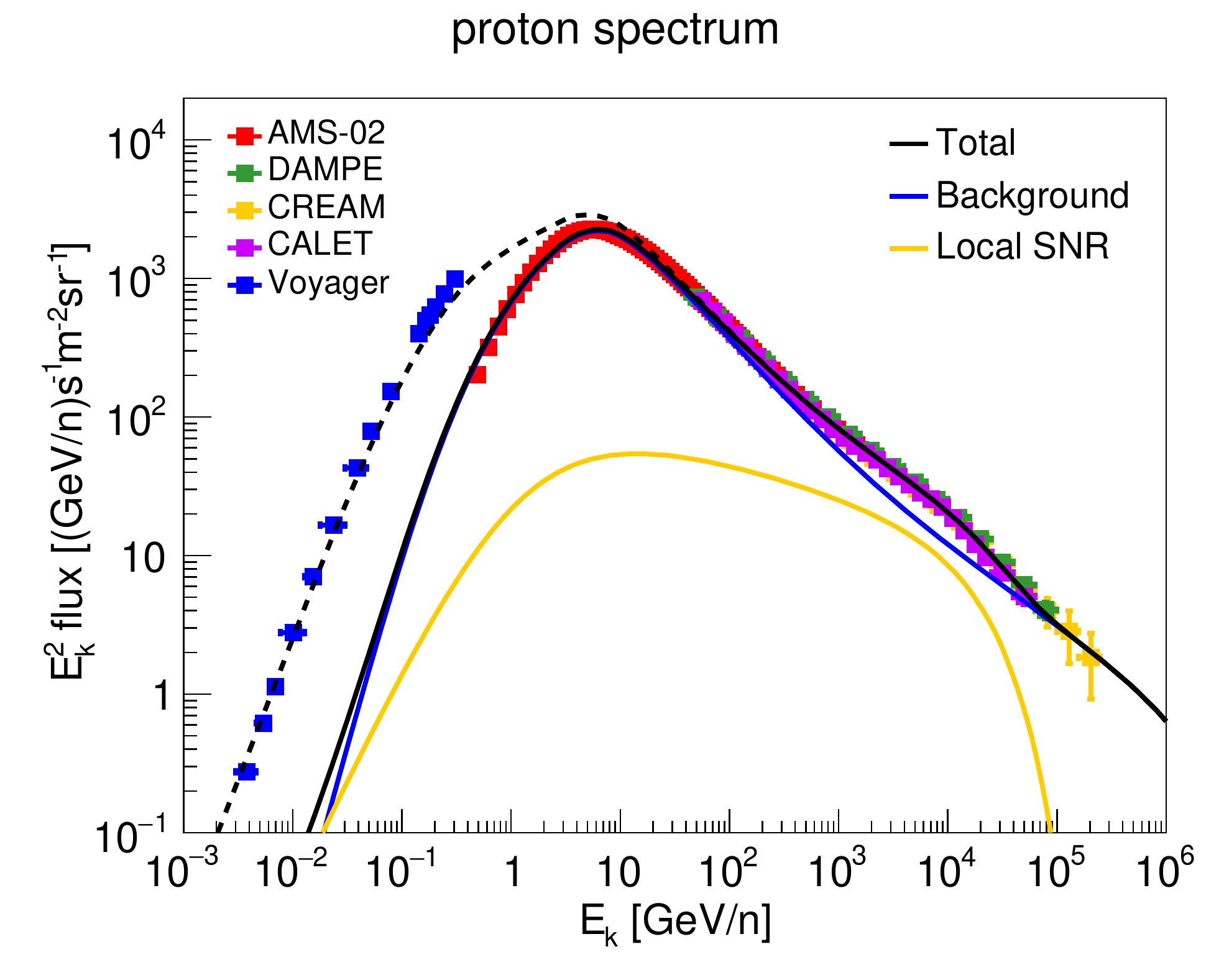}
\includegraphics[height=6.cm, angle=0]{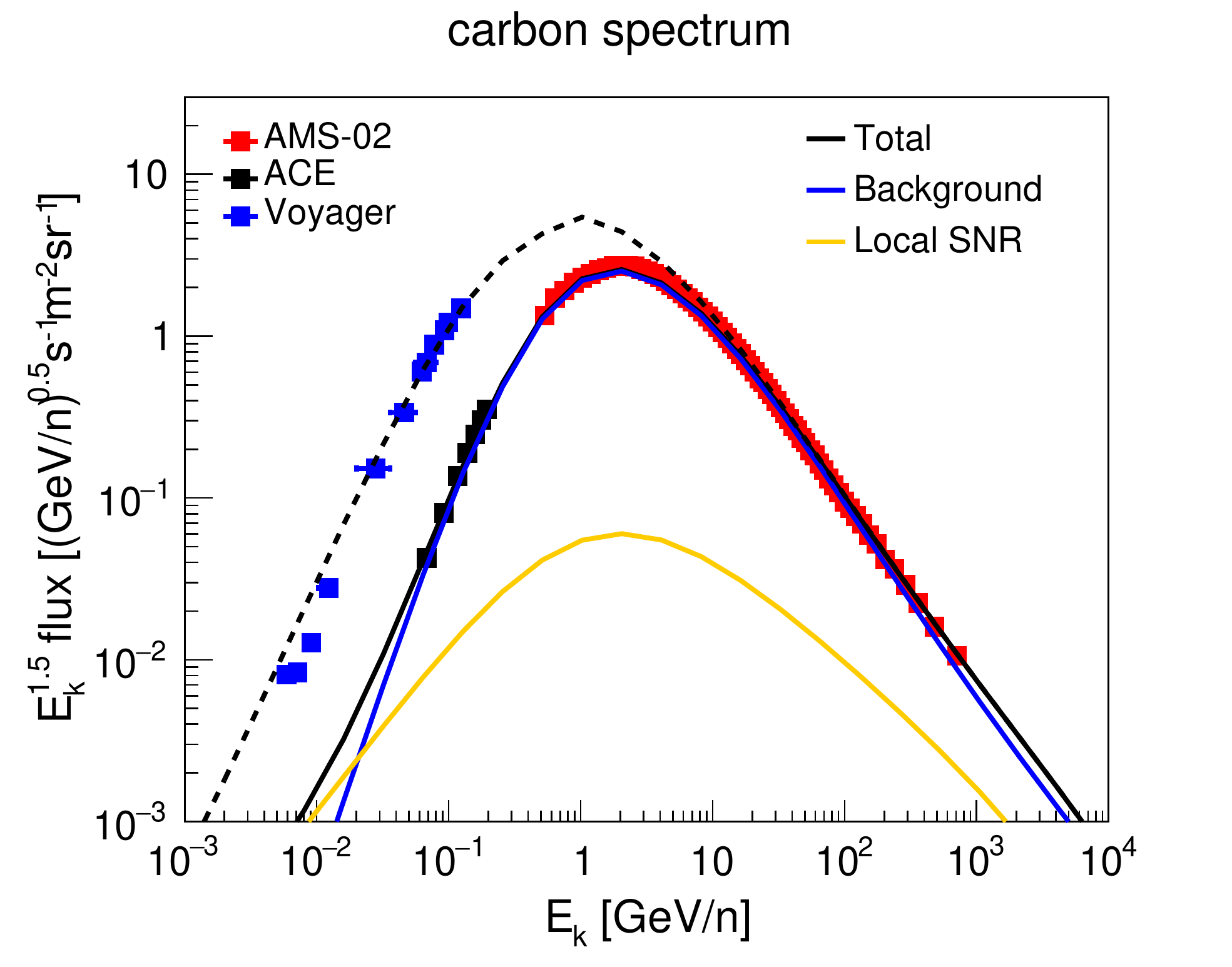}
\includegraphics[height=6.cm, angle=0]{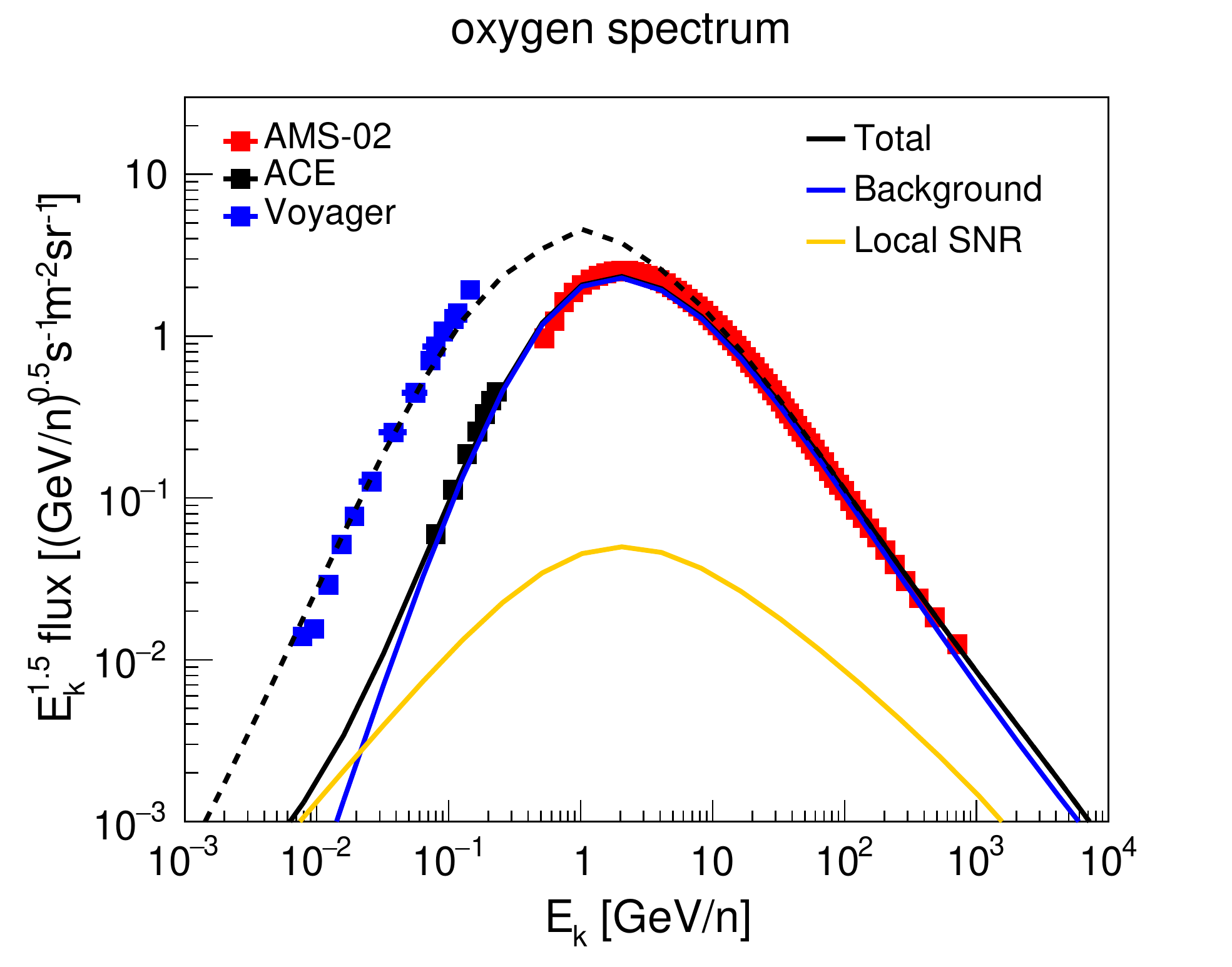}
\caption{
Calculated protons, carbon and oxygen spectra. The blue solid lines are the fluxes from background SNRs, and the yellow solid lines are the contribution from the local SNR. The black lines are the total fluxes. The black solid lines are the modulated ones. The data points are taken from the Voyager(blue) \citep{2016ApJ...831...18C},  ACE (black) \citep{2019SCPMA..6249511Y}, AMS-02 (red) \citep{2015PhRvL.114q1103A, 2018PhRvL.120b1101A}, DAMPE (green) \citep{2019SciA....5.3793A}, CALET (violet)\citep{2022PhRvL.129j1102A} and CREAM-II (yellow) \citep{2017ApJ...839....5Y} experiments.
}
\label{fig:pcoflux}
\end{figure*}

Fig. (\ref{fig:pcoflux}) displays the proton, carbon and oxygen spectra as a function of rigidity. Dashed lines are the local interstellar spectra (LIS), and solid lines are the modulated spectra. The yellow and blue solid lines are the calculated fluxes from the local SNR and the background sources, respectively. The black solid line represents the total flux. The fitted parameters of transport and the injection spectra are listed in Tab.\ref{table_1} and Tab.\ref{table-parm} respectively.
We add the Voyager data to compare with the low-energy local interstellar flux before solar modulation. We find that to account for the Voyager data in the diffusion-reacceleration scenario, a break in the primary injection spectra is necessary. For protons, carbon and oxygen, the break rigidities are $7.6$, $1.9$ and $1.9$ GV respectively. This is due to the fact that the break of the proton flux is at $\sim 10$ GV while for carbon and oxygen, the breaks are at a lower rigidity. And the power indexes change from $2.35$ to $1.41$ for protons and from $2.27$ to $1.00$ for carbon and oxygen. The corresponding modulation potentials are $650$, $660$ and $590$ MV. 

Above $200$ GV, the contribution from the local SNR is non-negligible. The hardening of proton, carbon and oxygen fluxes principally originate from the local SNR. To explain the softening of the proton spectrum at $\sim 20$ TeV observed by both CREAM and DAMPE experiments, $\mathcal R_{\rm c}$ is adjusted to $21$ TV. 

\begin{figure*}[!htb]
\centering
\includegraphics[height=6.cm, angle=0]{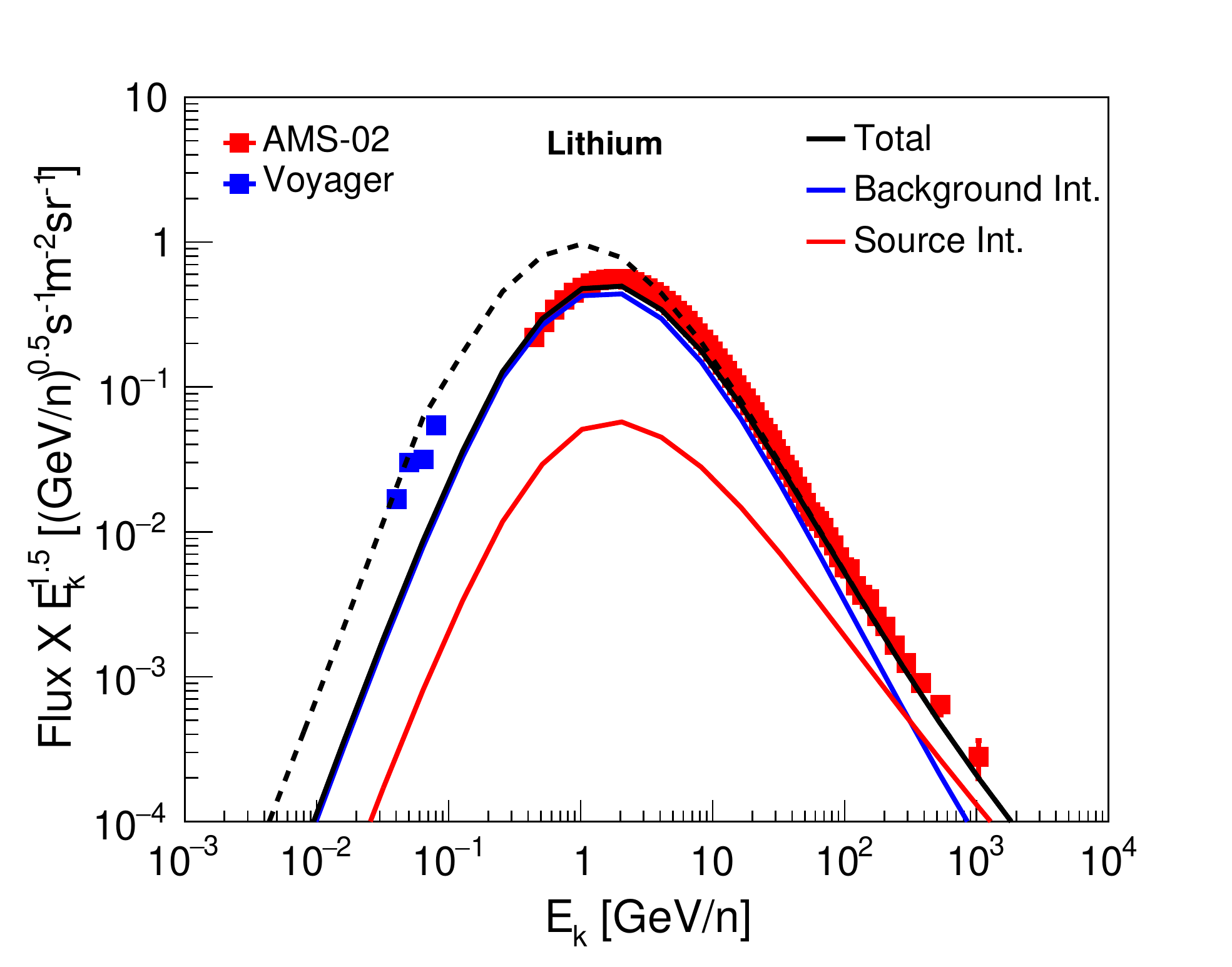}
\includegraphics[height=6.cm, angle=0]{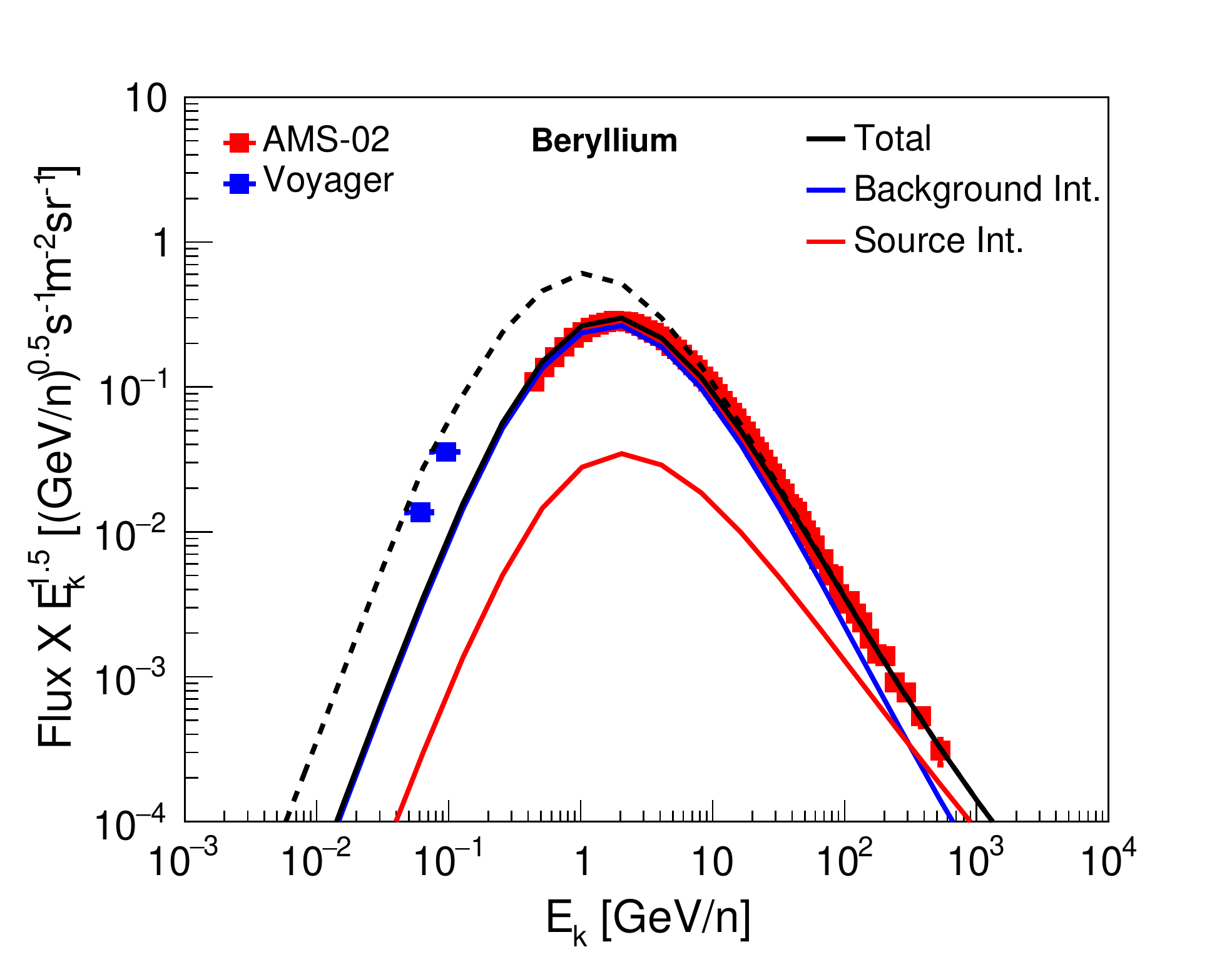}
\includegraphics[height=6.cm, angle=0]{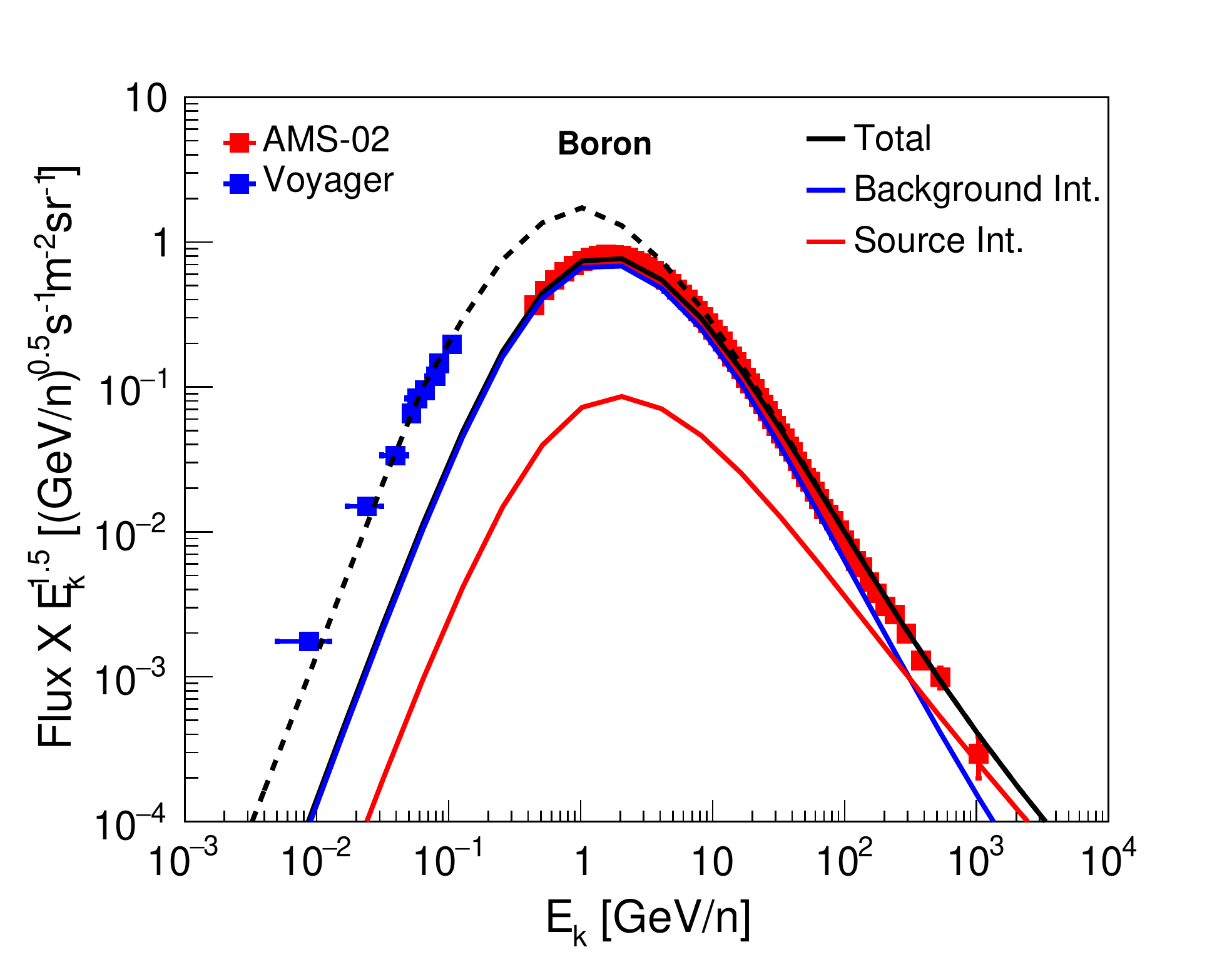}
\caption{
Calculated Lithium, Beryllium and Boron LIS (dashed black curve) and the modulated spectra at the Earth (solid black curve). The data points are taken from the Voyager\citep{2016ApJ...831...18C} and AMS-02 \citep{2018PhRvL.120b1101A} experiments.
}
\label{fig:bflux}
\end{figure*}

The calculated lithium, beryllium and boron fluxes are shown in Fig. \ref{fig:bflux}. The blue solid line corresponds to the secondary particles produced by primary particles interacting with the interstellar medium during propagation. The red solid line is lithium (beryllium, boron) generated near the source region. The power index after propagation is close to the background protons, and thus harder than the background boron.

\begin{figure*}
\centering
\includegraphics[height=6.cm, angle=0]{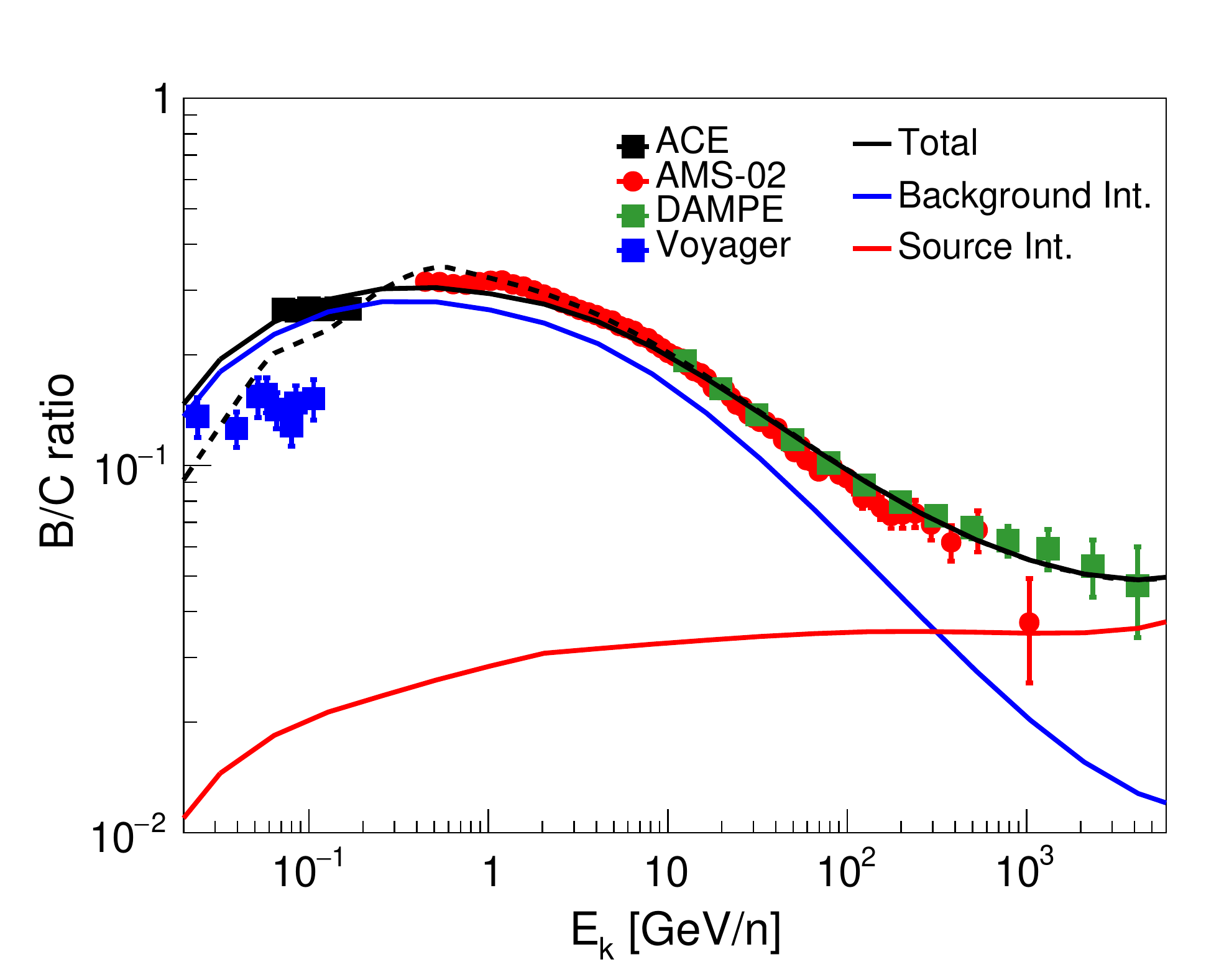}
\includegraphics[height=6.cm, angle=0]{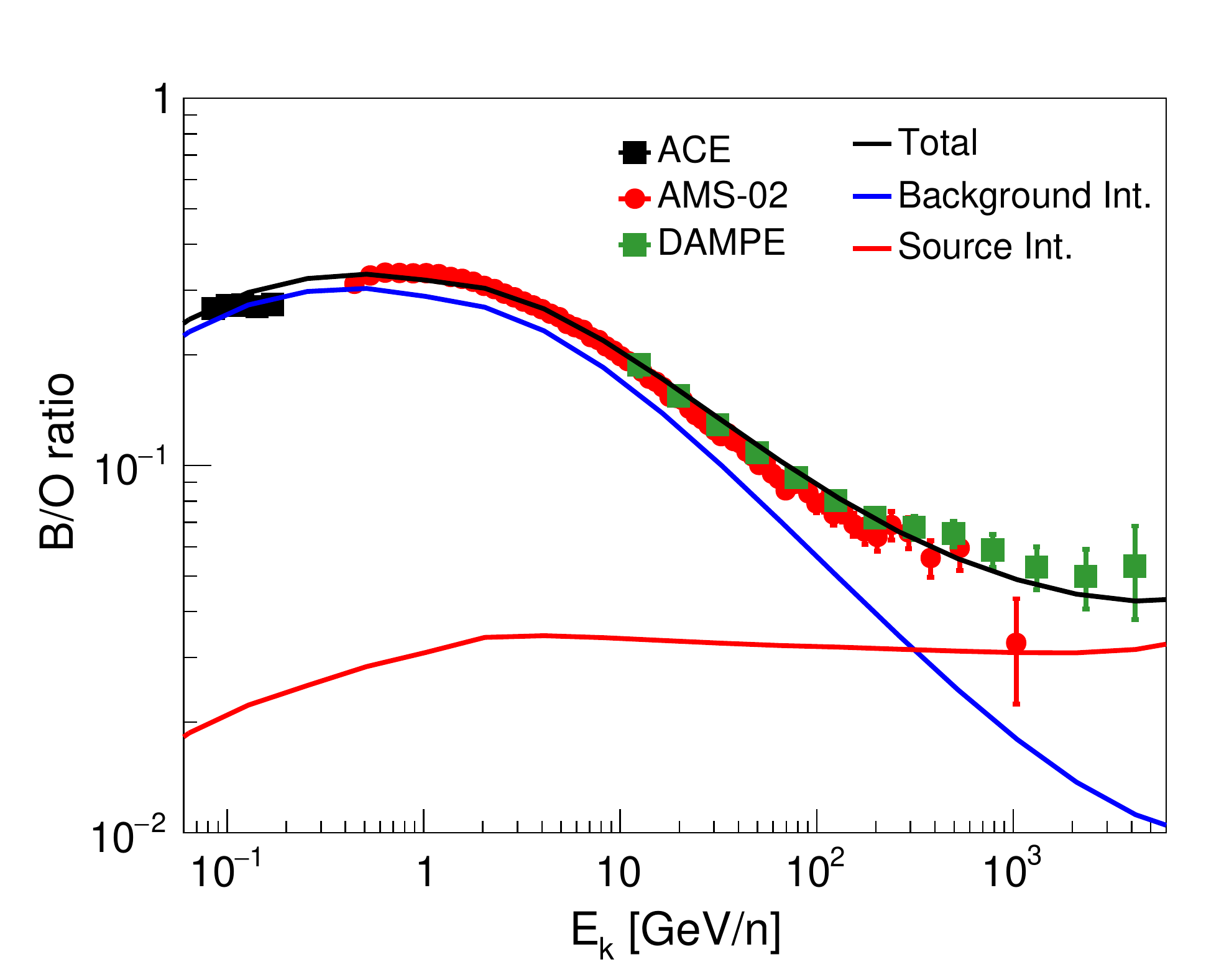}
\includegraphics[height=6.cm, angle=0]{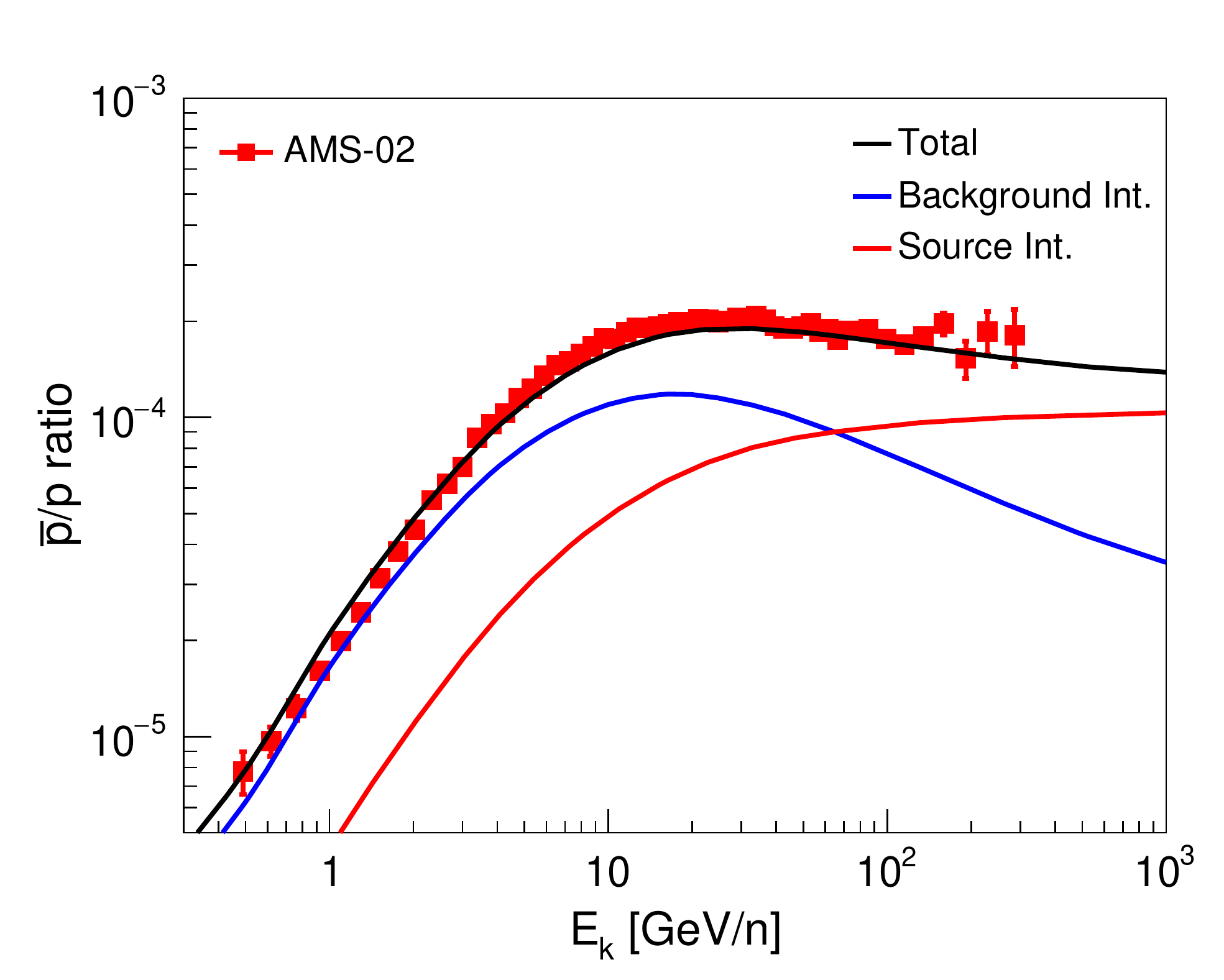}
\includegraphics[height=6.cm, angle=0]{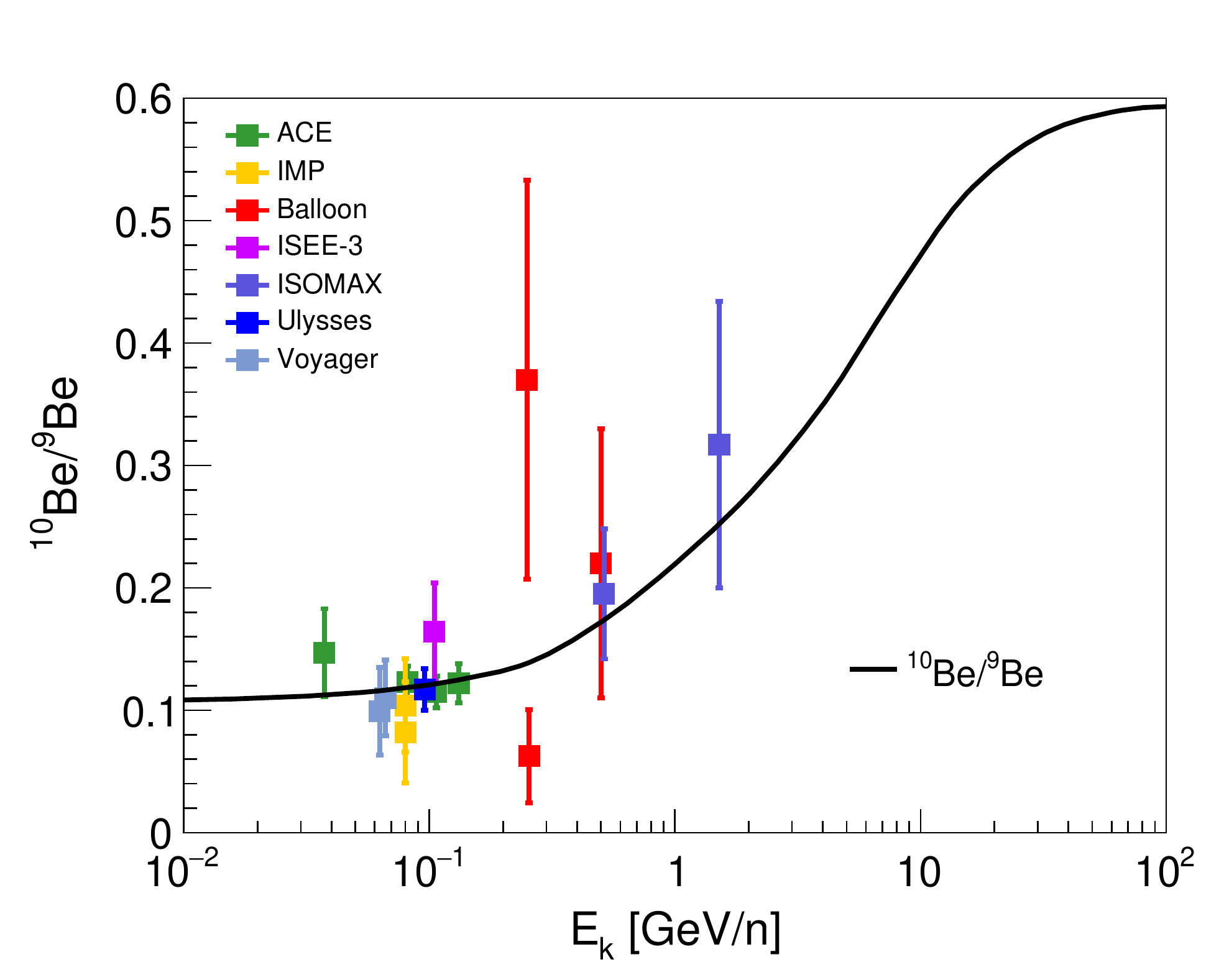}
\caption{
The ratios of boron-to-carbon, boron-to-oxygen, antiproton-to-proton and $^{10}$Be to $^{9}$Be. The data points are taken from the Voyager \citep{2016ApJ...831...18C}, ACE \citep{2019SCPMA..6249511Y,2001ApJ...563..768Y}, AMS-02 \citep{2018PhRvL.120b1101A}, DAMPE \citep{DAMPECOLLABORATION20222162}, IMP \citep{1981ICRC....2...72G}, Balloon \citep{1977ApJ...212..262H,1978ApJ...226..355B,1979ICRC....1..389W}, ISEE-3 \citep{1988SSRv...46..205S}, ISOMAX \citep{2004ApJ...611..892H} and Ulysses \citep{1998ApJ...501L..59C} experiments.
}
\label{fig:bcratio}
\end{figure*}

The Fig. (\ref{fig:bcratio}) shows calculations of the B/C, B/O, $\bar{p}/p$ and $^{10}$Be/$^{9}$Be ratios. The blue solid line is the ratio of boron ($\bar{p}$) generated during propagation over total carbon (oxygen, proton). 
The B/C ratio (i.e. red solid line), which is the flux of boron from nearby sources over total carbon flux, appears flatter. Since the carbon flux spectrum near the source region is harder compared to the propagated one, the corresponding boron spectrum generated near the source region is harder than that generated during propagation. Thus, the total B/C ratio is flatter above $1$ TeV and the same holds true for the B/O ratio. 
{Note that, in our model, the spectra of primary and secondary species are composed of two parts: the first part originates from the background sources and the fragmentations in the whole Galaxy, and the second part from a nearby source for primary CRs and the fragmentations around the source regions for secondary species. Considering the different origins of the components of primary and secondary species, it may be a coincidence that their spectra show apparent hardenings at similar rigidities. On the other hand, it should be noticed that the hardening rigidity for the spectra of different CR species may not be exactly uniform. The fittings to the AMS-02 data showed that the break rigidity has a diversity (see e.g., \cite{2022FrASS...944225N,2022ApJ...932...37N}). Combining the higher energy measurements by DAMPE and CALET also gives higher break rigidity than 200 GV \cite{2019SciA....5.3793A,2021PhRvL.126t1102A,2020PhRvL.125y1102A}. Therefore the similar values of the break in the primary spectra and in the secondary/primary ratios at hundreds of GV may just be a coincidence.
}
As for the $\bar{p}/p$ ratio, the antiproton spectrum from source regions is harder than that produced during propagation. The $\bar{p}/p$ ratio becomes flatter above $10$ GeV, so the total $\bar{p}/p$ ratio is energy independent above $10$ GeV. Similar to other ratios, our model calculation for $^{10}Be/^9Be$ is consistent with the measurements. 

\begin{figure*}
\centering
\includegraphics[height=14.cm, angle=0]{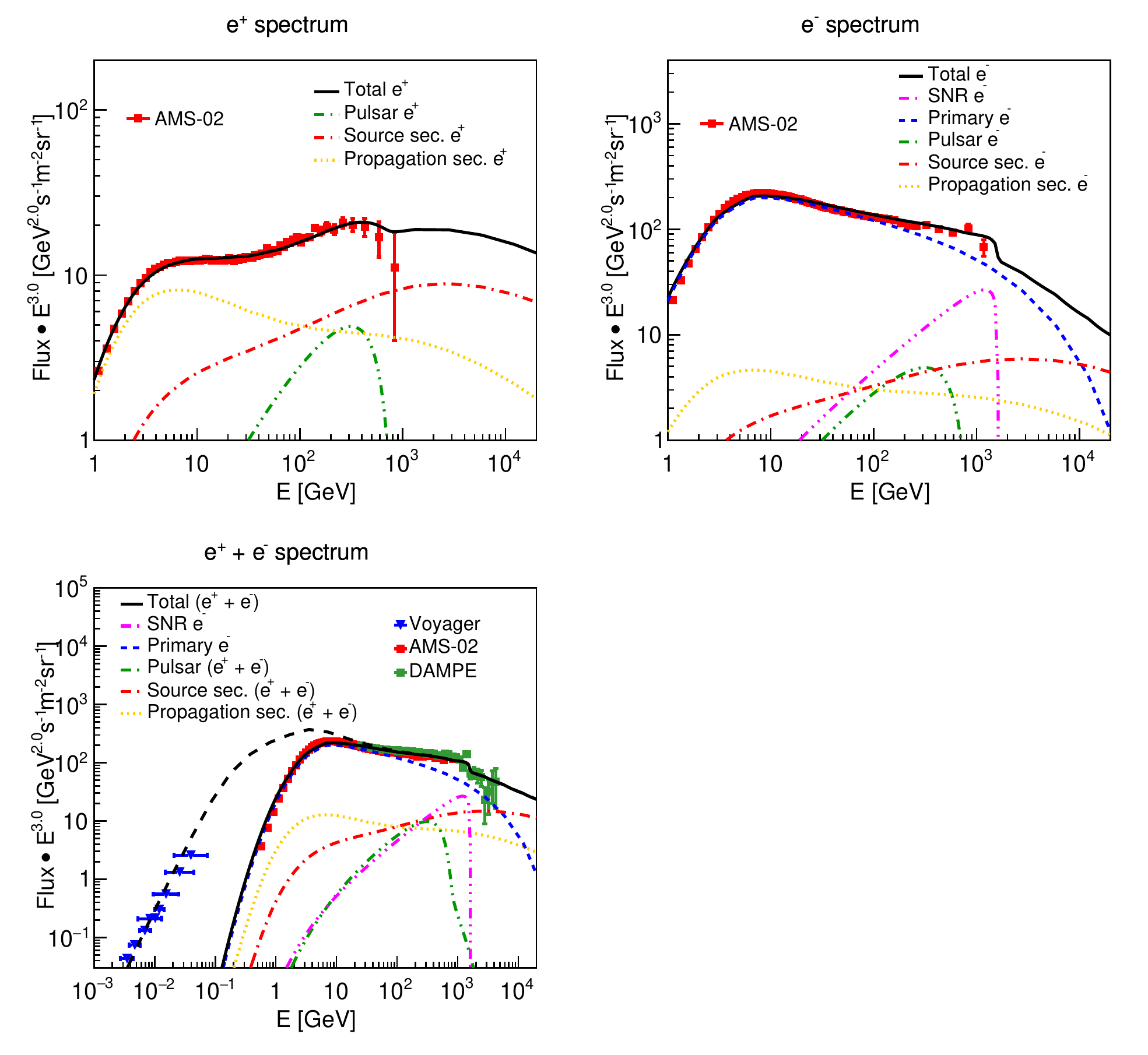}
\caption{
Calculated positron and electron energy spectra. The top panels are the positron and electron spectra, while the bottom panel is the all-electron spectrum. The black solid lines are the total fluxes. The data points are taken from the Voyager \citep{2016ApJ...831...18C}, AMS-02 \citep{2019PhRvL.122d1102A, 2019PhRvL.122j1101A}, and DAMPE \citep{2017Natur.552...63D} experiments.
}
\label{fig:elec}
\end{figure*}

Finally, we discuss the results of positrons and electrons, as is shown in Fig. (\ref{fig:elec}). The positrons have three origins: secondary ones from the contributions of CRs during propagation in the Milky Way and around the accelerators, and primary ones contributed by the local pulsars. As for CR electrons, in addition to the identical components as positrons, there are also primary components from both the background sources and the local SNR. The parameters of electrons and positrons are also given in Table \ref{table-parm}. As we can see, for the positrons, the source region component dominates over those from propagation and local pulsars. For electrons and electrons plus positrons, below $1$ TeV, the propagation contribution is dominant, but beyond that energy, the source contribution becomes important; above tens of TeV, the propagation contribution damps exponentially, explaining the break in the all-electron spectrum at TeV energies. Meanwhile, the total flux still has a significant change compared to that expected from propagation only. This could be tested by further precise measurements.

\section{Conclusion}

The CR observations have entered an era of precise measurements. More and more experiments demonstrate that the measurements are far from the expectations of the conventional CR model. More recently, the DAMPE experiment has reported that the boron-to-carbon ratio exhibits a significant excess above $200$ GV. This indicates a need of modification of either the acceleration model or the propagation model, if not both.

In this work, we propose that the excesses of secondary CRs originate from the hadronic interactions between the freshly accelerated cosmic rays and the medium, as CRs experience a slower diffusion near the sources. This process has been neglected in the studies of the propagation model. In comparison with the secondaries generated during propagation, the secondary CR flux generated in the source regions is harder and extends to TeV energies. Therefore, the ratio anomalies, i.e., boron-to-carbon, boron-to-oxygen, and antiproton-to-proton ratios could be naturally accounted for. 

What is even more interesting is that above TeV energies, the electron and positron spectra give rise to extra hardening. This is caused by the positrons and electrons generated near the sources. We hope this prediction could be tested by the future experiments, for example HERD.


\acknowledgments
This work is supported by the National Key Research and Development Program of China
(Nos. 2018YFA0404203), the National Natural Science Foundation of China (Nos. 12220101003, 12275279, U2031110) and CAS Project for Young Scientists in Basic Research (No. YSBR-061).

\bibliographystyle{unsrt_update}
\bibliography{ref}

\end{document}